# Energy-dependent normal and unusually large inverse chlorine kinetic isotope effects of simple chlorohydrocarbons in collision-induced dissociation by gas chromatography-tandem mass spectrometry

**Caiming Tang[a,*], Jianhua Tan[b], Peilin Zhang[b], Yujuan Fan[a,c], Zhiqiang Yu[a], Xianzhi Peng[a,*]**

[a] *State Key Laboratory of Organic Geochemistry, Guangzhou Institute of Geochemistry, Chinese Academy of Sciences, Guangzhou 510640, China*

[b] *Guangzhou Quality Supervision and Testing Institute, Guangzhou, 510110, China*

[c] *University of Chinese Academy of Sciences, Beijing 100049, China*

*Corresponding Authors.

Tel: +86-020-85291489; E-mail: CaimingTang@gig.ac.cn (C. Tang).

Tel: +86-020-85290009; Fax: +86-020-85290009. E-mail: pengx@gig.ac.cn (X. Peng).

# ABSTRACT

Kinetic isotope effects (KIEs) taking place in mass spectrometry (MS) can provide in-depth insights into the fragmental behaviors of compounds in MS. Yet the mechanisms of KIEs in collision-induced dissociation (CID) in tandem MS are unclear, and information about chlorine KIEs (Cl-KIEs) of organochlorines in MS is particularly scarce. This study investigated the Cl-KIEs of dichloromethane, trichloroethylene and tetrachloroethylene during CID using gas chromatography-electron ionization triple-quadrupole tandem MS. Cl-KIEs were measured with MS signal intensities, and their validity was confirmed in terms of chromatograms, crosstalk effects and background subtraction influences. All the organochlorines presented large inverse Cl-KIEs, showing the largest values of 0.492, 0.910 and 0.892 at the highest collision energy for dichloromethane, trichloroethylene and tetrachloroethylene, respectively. For dichloromethane, both intra-ion and inter-ion Cl-KIEs were studied, within the ranges of 0.492-1.020 and 0.614-1.026, respectively, showing both normal and inverse Cl-KIEs depending on collision energies. The observed Cl-KIEs generally declined with the increasing collision energies from 0-60 eV, but were inferred to be independent of MS signal intensities. The normal Cl-KIEs were explained with the quasi-equilibrium theory and zero point energy theory, and the inverse Cl-KIEs were interpreted with isotope-competitive reactions relevant to transition state theory. The Cl-KIEs are dominated by critical energies at low internal energies, while controlled by rotational barriers (or looseness/tightness of transition states) at high internal energies. It is concluded that the Cl-KIEs may depend on critical energies, bond strengths, available internal energies, and transition state looseness/tightness. The findings of this study yield new insights into the fundamentals of Cl-KIEs of organochlorines during CID, and may be conducive to elucidating the mechanisms of KIEs in collision-induced and photo-induced reactions in the actual world.

## Keywords:

Chlorine kinetic isotope effects; Chlorohydrocarbons; Collision-induced dissociation; Electron ionization tandem mass spectrometry; Collision energy; Isotope-competitive reactions

# 1. Introduction

Kinetic isotope effects (KIEs) occurring in mass spectrometry (MS) have been extensively studied for quite a long time in terms of observations, theories, mechanisms and applications [1-4], which may provide in-depth insights into fragmental behaviors of compounds in MS. Large KIEs can take place during fragmentation in ionization source, metastable-ion decomposition in mass analyzer, and collision-induced dissociation (CID) in tandem MS (MS/MS) [5-7]. By means of MS, KIEs can be readily determined and applied to probing reaction mechanisms of molecules and ions in gaseous phase [8-10]. So far, most studies relevant to on-MS KIEs focused on hydrogen/deuterium (H/D) KIEs [5,11-13], whereas a few concerned heavy-atom KIEs such as chlorine KIEs (Cl-KIEs) [14-16].

Cl-KIEs can take place in electron ionization-MS (EI-MS) during both ionization and metastable-ion dissociation [17]. In addition, some studies have reported the Cl-KIEs of some mono-chlorinated organic compounds [18], chloroform [19], tetrachloromethane [20], and chloride adducts of some chlorinated/non-chlorinated organic compounds [14,16,19] taking place during CID in MS/MS. Both normal and inverse Cl-KIEs were observed in these studies with very large magnitudes in comparison with those during reactions in solution. Some mechanistic interpretations have been proposed for the Cl-KIEs present in CID. Zakett et al. pointed out that the magnitudes of Cl-KIEs of organochlorines were dependent on the internal energies and structures of ions [18], and Lehman et al. indicated that the Cl-KIEs of tetrachloromethane simply relied on a function of the masses of ions instead of detection parameters including detector slit width and accelerating voltage [20]. Petersen et al. found large normal Cl-KIEs of chloroform during CID, and concluded that the dechlorination reaction proceeded via two different pathways and the Cl-KIEs were symmetry-induced [19]. Extremely large Cl-KIEs of $CHCl_4^-$ during CID were observed and inferred to be owing to the near-threshold centrifugal effect [14], while Petersen et al. concluded the Cl-KIEs were caused by symmetry-induced degeneracy of vibrational modes [19]. Augusti et al. observed large normal and inverse Cl-KIEs of chloride adducts of aliphatic alcohols, benzaldehyde, and 2,4-pentanedione during CID [16], and deduced that the Cl-KIEs were triggered by differences of zero-point energies (ΔZPEs) between isotopomers, which was in agreement with the conclusions of Gozzo et al [15]. However, Lehman et al. indicated that KIEs in CID could not ascribe to ΔZPEs based on the observed KIEs of diatomic ions [20]. Thus, it can be seen that the fundamentals of KIEs in CID are still equivocal at present. Furthermore, the relationships between Cl-KIEs of organochlorines in CID and internal energies have not been unraveled yet, let alone whether both normal and inverse Cl-KIEs of an organochlorine can occur along with the variation of internal energies that depend on collision energies (CEs).

Chlorohydrocarbons such as dichloromethane (DCM), trichloroethylene (TCE) and tetrachloroethylene (PCE) are important pollutants in the environment [21,22], particularly in the atmosphere and groundwater [23-26]. Atmospheric chlorohydrocarbons may be subjected to dechlorination reactions caused by collision with high-energy charged particles of solar wind like the formation of aurora [27], or triggered by excitation of solar radiation and cosmic rays. During these dechlorination reactions, Cl-KIEs may be present, thereby affecting the chlorine isotope ratios and isotopologue compositions of chlorohydrocarbons and their products [14,28,29]. CID in EI-MS/MS may be an alternative approach to simulate the ion-molecule collision-induced dechlorination reactions and obtain Cl-KIEs to provide clues in mechanistic elucidation for collision-induced and photo-induced dechlorination reactions of atmospheric chlorohydrocarbons.

Hence, we conducted a systematic study to investigate the Cl-KIEs of chlorohydrocarbons in CID by gas chromatography electron ionization triple-quadrupole tandem mass spectrometry (GC-MS/MS) using DCM, TCE and PCE as model compounds. By changing CEs within a relatively large range, the relationships between Cl-KIEs and internal energies were revealed. Theories in terms of isotope-competitive reactions were applied to interpreting mechanisms of the internal energy-dependent normal and large inverse Cl-KIEs of DCM. This study provides a clear and new understanding on the basic mechanisms of Cl-KIEs of organochlorines during CID, and may shed light on the mechanisms of Cl-KIEs of organochlorines in collision-induced and photo-induced degradations in the real world.

## 2. Materials and methods

### 2.1. Chemicals and materials

Dichloromethane (DCM, purity: 99.8%) was purchased from BCR International Trading Co., Ltd (California, USA), and trichloroethylene (TCE, purity: 99.0%) along with tetrachloroethylene (PCE, purity: 99.0%) were bought from Dr. Ehrenstorfer (Augsburg, Germany). Solvent n-hexane was obtained from Merck Corp. (Darmstadt, Germany). All the chemicals were of chromatographic grade and used without further purification. The standards of DCM, TCE and PCE were accurately weighed and dissolved in n-hexane to prepare stock solutions at 1.0 mg/mL. The stock solutions were further diluted with n-hexane to prepare working solutions at 100.0 µg/mL. All the standard solutions were stored at −20 °C condition prior to instrumental analysis.

### 2.2. GC-MS/MS analysis

The GC-MS/MS system consisted of an Agilent 7890B gas chromatograph and an Agilent 7000D triple quadrupole mass spectrometer (Agilent Technologies, Palo Alto, CA, USA). The chromatographic separation was conducted with an Agilent DB-INNOWAX capillary column (30 m × 0.25 mm, 0.25 µm thickness, Agilent Technologies). The temperature programs and other GC parameters are detailed in Table S-1. The instrument control and data processing were performed with MassHunter Workstation (Agilent Technologies). The working solutions were directly injected onto the GC-MS/MS system with the injection volume of 1 µL, and splitless injection mode was adopted. At each instrumental condition, six replicated measurements were carried out.

The MS/MS working conditions and parameters for DCM are detailed as follows:

positive EI source was used; EI energy was 45 eV; ionization source temperature was 250 °C; multiple reaction monitoring (MRM) mode was used for data acquisition; dwell time of each MRM transition was 25 ms; mass resolution was set at unit level (0.7 u) for both MS1 and MS2; collision gas was $N_2$; CEs were 0, 0.1, 0.2, 0.5, 1, 2, 3, 4, 5, 6, 8, 10, 12, 15, 20, 30, 40, 45, 50 and 60 eV. The working conditions and parameters of the MS/MS for TCE and PCE are generally similar to those for DCM. Some specific MS/MS parameters for TCE and PCE are detailed as the following: the ionization source was maintained at 280 °C; dwell time of each MRM transition was 30 ms; CEs were 0, 0.1, 0.2, 0.5, 1, 2, 5, 10, 15, 20, 30, 40, 45, 50, and 60 eV; solvent delay time was 3 min; scan window for TCE was 3-4.7 min and that for PCE was 4.7-12.7 min. The detailed information related to the MRM transitions is provided in Table S-2, and the representative MRM chromatograms are shown in Figure 1.

*2.3. Data processing*

The intra-ion Cl-KIE of DCM was calculated with

$$KIE_{Intra} = I_{86.0 \to 51.0} \big/ I_{86.0 \to 49.0} \tag{1}$$

where $I$ denotes the MS signal intensities corresponding to the respective MRM transitions indicated by the subscripts, i.e., $m/z$ 86.0→51.0 and 86.0→49.0 (similarly hereinafter). In addition, the inter-ion Cl-KIE of DCM was calculated as:

$$KIE_{Inter} = I_{84.0 \to 49.0} \times R_{Cl}^{2} \big/ I_{88.0 \to 51.0} \tag{2}$$

where $R_{Cl}$ is the measured chlorine isotope ratio of DCM calculated as follows:

$$R_{Cl} = \frac{I_{88.0 \to 88.0} \times 2 + I_{88.0 \to 51.0} + I_{86.0 \to 86.0} + I_{86.0 \to 51.0}}{I_{84.0 \to 84.0} \times 2 + I_{84.0 \to 49.0} + I_{86.0 \to 86.0} + I_{86.0 \to 49.0}} \tag{3}$$

which is derived from previously reported schemes of isotope-ratio calculation [17,30,31]. The calculation schemes for the intra-ion Cl-KIEs of TCE and PCE, similar to that of DCM, are detailed in the Supplementary data. All these calculation schemes originate from the literature [1]. The total MS signal intensity of each MRM transition within the retention time range of each chromatographic peak was exported from the corresponding raw data file and applied to Cl-KIE calculation. Prior to exporting MS signal intensities, background subtraction was conducted by subtracting the baseline signal intensities at both ends of each peak from the intensity of the peak. Data from six replicated injections were used to calculate the average Cl-KIEs and MS intensities along with their standard deviations ($1\sigma$).

### 2.4. Statistical analysis

Statistical analysis was conducted with SPSS Statistics 19.0 (IBM Inc., Armonk, USA) and Origin 9 (OriginLab Corp., Northampton, USA). Paired-samples T test, performed with SPSS, were applied to examining the differences of Cl-KIEs by determining p-values (2-tailed) with alpha of 0.01 as the threshold value for significance. If a p-value is $\leq 0.01$, the null hypothesis (e.g., no difference between two Cl-KIEs or two types of Cl-KIEs) is rejected, indicating an indeed existent significant difference. Linear and non-linear regressions, implemented by Origin, were employed to reveal the relationships between experimental data and parameters/factors of interest.

# 3. Results and discussion

## 3.1. Validity of measured Cl-KIEs

In this study, we aimed to unravel the relationships between Cl-KIEs and internal energies during CID in MS/MS. The measured Cl-KIEs were critical to implement the revelation of the relationships, and therefore it was crucial to confirm their validity prior to application. As shown in Figure 1, the MRM transition of $m/z$ 86.0→49.0 showed apparently higher abundances than $m/z$ 86.0→51.0 when CEs were higher than 30 eV, visually indicating differentiable extents of the two branch reactions.

Crosstalk during MRM might be a factor capable of affecting the validity of measured Cl-KIEs. For DCM, seven MRM transitions were monitored (Table S-2). Both the precursor ions of $m/z$ 84.0 and 86.0 can give rise to the product ion of $m/z$ 49.0, and both the precursor ions of $m/z$ 86.0 and 88.0 can yield the product ion of $m/z$ 51.0. As a consequence, if crosstalk presents, the precursor ion of $m/z$ 84.0 could influence the monitoring of the MRM transition of $m/z$ 86.0→49.0 by enhancing the MS signal intensity. On the other hand, the precursor ion of $m/z$ 86.0 could enhance the MS signal intensity of the transition of $m/z$ 88.0→51.0. We accordingly conducted an experiment by monitoring the transitions of $m/z$ 86.0→49.0 and 86.0→51.0 exclusively, and another experiment monitoring the transitions of $m/z$ 84.0→49.0 and 88.0→51.0 only. As shown in Figure 2, Cl-KIEs obtained by monitoring all MRM transitions (combined scan) are very close to those by monitoring the transitions used for determining intra-ion and inter-ion Cl-KIEs separately (separated scan). This result demonstrates that the crosstalk might not be present in this study or could not impact the determination of Cl-KIEs.

In this study, background subtraction was carried out before exporting MS signal intensity, which might be another issue affecting the validity of measured Cl-KIEs. Background subtraction might take away slight real signal intensities from MRM transitions. And transitions with lower theoretical relative abundances were speculated to be affected by background subtraction more significantly in contrast to transitions with higher relative abundances. If so, the inter-ion Cl-KIEs obtained with background subtraction should be higher than those without background subtraction. However, as illustrated in Figure 2, the inter-ion Cl-KIEs obtained with and without subtraction are almost identical (p = 0.9), which demonstrates that the background subtraction had no effect on the measured Cl-KIEs. Ultimately, the validity of measured KIEs can be accordingly confirmed.

### 3.2. Measured Cl-KIEs

#### 3.2.1. Relationships between Cl-KIEs and collision energies

Both intra-ion and inter-ion Cl-KIEs of DCM during CID in GC-EI-MS/MS were investigated in this study. The intra-ion Cl-KIEs were derived from the reactions of $CH_2{}^{35}Cl{}^{37}Cl^{•+} \rightarrow CH_2{}^{35}Cl^+$ ($m/z$ 86.0→49.0) vs. $CH_2{}^{35}Cl{}^{37}Cl^{•+} \rightarrow CH_2{}^{37}Cl^+$ ($m/z$ 86.0→51.0); and the inter-ion Cl-KIEs were derived from the reactions of $CH_2{}^{35}Cl_2{}^{•+} \rightarrow CH_2{}^{35}Cl^+$ ($m/z$ 84.0→49.0) vs. $CH_2{}^{37}Cl_2{}^{•+} \rightarrow CH_2{}^{37}Cl^+$ ($m/z$ 88.0→51.0). The rate constants corresponding to the reactions by losing $^{35}Cl$ or $^{37}Cl$ atoms are denoted as $k_{35Cl}$ and $k_{37Cl}$, respectively and the Cl-KIE is $k_{35Cl} / k_{37Cl}$ (i.e., Cl-KIE = $k_{35Cl} / k_{37Cl}$). As shown in Figure 3, both the intra-ion and inter-ion Cl-KIEs generally declined as the CEs increased from 0 to 60 eV. The highest intra-ion and inter-ion KIEs were 1.020±0.001 and 1.026±0.002, respectively and the lowest were 0.492±0.008 and 0.614±0.013, respectively (Table S-4). We divide the CEs set for detecting DCM into two regions, viz. a low energy region (0-5 eV) and a high energy region (6-60 eV). In the low energy region, the main intra-ion Cl-KIEs (at 0-2 eV) and the major inter-ion Cl-KIEs were significantly higher than unity where $k_{35Cl}$ is equal to $k_{37Cl}$ (p $\leq$ 0.003, Figure 3a). Whereas all the Cl-KIEs in the high energy region were significantly lower than unity (p $\leq$ 0.001, Figure 3b). In the entire low energy region and in 20-60 eV, all the intra-ion Cl-KIEs were lower than the inter-ion Cl-KIEs at individual CEs (p $\leq$ 0.004), whereas in 6-15 eV the differences between the two types of Cl-KIEs were insignificant (p $\geq$ 0.05, Table S-4), except at the CE of 8 eV at which the intra-ion Cl-KIE was slightly higher than the inter-ion Cl-KIE (p = 0.004). The correlations between Cl-KIEs and CEs can be fitted well with exponential functions ($R^2 \geq$ 0.959, Figure 3). It is noteworthy that the fitted curves were concave in the low energy region but convex in the high energy region, manifesting a decelerating decline of Cl-KIEs as the CEs from 0 to 5 eV and an accelerating decline with the CEs from 6 to 60 eV. This observation may imply different mechanisms of Cl-KIEs in the low and the high energy regions.

Since we could not obtain the exact relative abundances of the precursor ions $CH_2^{35}Cl_2^{\bullet+}$ and $CH_2^{37}Cl_2^{\bullet+}$ which transformed into the product ions $CH_2^{35}Cl^+$ and $CH_2^{37}Cl^+$, respectively, we thus determined the overall chlorine isotope ratios to adjust the measured MS signal intensity ratios of $m/z$ 84.0→49.0 to 88.0→51.0 for attaining the rough inter-ion Cl-KIEs (Eqs 2 and 3). As shown in Figure S-1, the raw intensity ratios between $m/z$ 84.0→49.0 and 88.0→51.0 generally declined along with the CEs from 0 to 60 eV, which coincides well with the correlations between the inter-ion KIEs and CEs as illustrated in Figure 3. Therefore, the measured inter-ion Cl-KIEs can correctly reflect the relationships between the exact inter-ion Cl-KIEs and CEs, even though the measured inter-ion Cl-KIEs were approximate values.

In addition to DCM, TCE and PCE were also used as model compounds to implement the investigation of intra-ion Cl-KIEs during CID in GC-EI-MS/MS. Due to the low abundances of the isotopologues whose chlorine atoms are merely $^{37}$Cl, the inter-ion Cl-KIEs were not studied. In addition, because TCE and PCE have more than two Cl atoms, the degrees of freedom for losing Cl atoms should be taken into account (Eqs S1, S2 and S3). As shown in Figure 4, the measured Cl-KIEs linearly decreased as the CEs varied from 0 to 60 eV. The highest Cl-KIEs derived from the reactions of $m/z$ 131.9→94.9 vs. 131.9→96.9, and $m/z$ 133.9→96.9 vs. 133.9→98.9 for TCE were 0.999±0.003 and 0.992±0.007, respectively, and the lowest were 0.910±0.005 and 0.926±0.009, respectively (Figure 4a and Table S-6). As for PCE, the highest Cl-KIEs derived from the reactions of $m/z$ 165.9→128.9 vs. 165.9→130.9 and $m/z$ 167.9→130.9 vs. 167.9→132.9 were 0.982±0.005, and 0.986±0.004, respectively, and the lowest were 0.905±0.010 and 0.892±0.005, respectively (Figure 4b and Table S-6). The fitted lines for the correlations between Cl-KIEs and CEs are very similar, showing indistinguishable Cl-KIEs of each compound at individual CEs ($p \geq 0.04$). Besides the Cl-KIEs of PCE derived from $m/z$ 167.9→130.9 vs. 167.9→132.9 by losing one Cl atom, the Cl-KIEs derived from $m/z$ 167.9→93.9 vs. 167.9→97.9 by losing two Cl atoms were also investigated with the CEs from 20 to 60 eV. As illustrated in Figure 4c, the Cl-KIEs derived from the reaction by losing one Cl were significantly lower than those from the reaction by losing two Cl atoms ($p = 5 \times 10^{-6}$). It is worth of noting that the Cl-KIE derived from the reaction losing two Cl atoms at CE of 20 eV exceeded unity with statistical significance (1.016±0.006, $p = 0.01$), and that at 30 eV was very close to unity (0.999±0.008, $p = 0.2$). Nevertheless, other Cl-KIEs of PCE were all under unity.

*3.2.2. Relationships between Cl-KIEs and MS intensities*

For revealing the in-depth mechanisms of Cl-KIEs during CID, correlations between Cl-KIEs and MS signal intensities were investigated. As shown in Figure 5, in the low energy region, the Cl-KIEs of DCM approximately linearly decreased as the intensities increased. However, in the high energy region the Cl-KIEs showed a decelerating increase along with the increasing intensities. This result indicates that there is no necessary connection between Cl-KIEs and MS signal intensities. In addition, we plotted the MS signal intensities vs. CEs, and find the intensities increased as the CEs varied from 0 to 5 eV and decreased with the CEs from 6 to 60 eV (Figure S-2). This observation indicates that the MS signal intensities are CE-dependent, just as the Cl-KIEs, but their CE-dependent variation modes are different (Figure 3 and Figure S-2). Similar conclusions can also be drawn from the data of TCE and PCE (Figure S-3 and Figure S-4). Accordingly, we can further infer that the Cl-KIEs occurring during CID are independent of injection concentrations and volumes of compounds and relative abundances of isotopologues.

### 3.3. Mechanistic interpretation

#### 3.3.1. Cl-KIEs during further dechlorination of product ions

In this study, because the lost Cl atoms during CID cannot be detected, we thus detected the specific product ions and used their signal intensities to calculate Cl-KIEs. These product ions contains at least one Cl atom, which might be subjected to further dechlorination during CID or metastable-ion dissociation. Cl-KIEs are anticipated to be present in the further dechlorination reactions, therefore maybe affecting the measured Cl-KIEs. As a result, the correlations between Cl-KIEs and CEs attained in this study need scrutiny in terms of the influence from further dechlorination reactions of product ions. We use DCM as an exemplary compound to elucidate the influence, and take the intra-ion Cl-KIEs into consideration at first. The intra-ion Cl-KIEs were derived from the reactions $CH_2{}^{35}Cl^{37}Cl^{\bullet+} \rightarrow CH_2{}^{37}Cl^+$ and $CH_2{}^{35}Cl^{37}Cl^{\bullet+} \rightarrow CH_2{}^{35}Cl^+$. As generally expected, a reaction losing a $^{35}Cl$ atom is faster than that losing a $^{37}Cl$ atom, i.e., $k_{35Cl} > k_{37Cl}$ or Cl-KIE > 1. We hypothesize that the reaction of $CH_2Cl_2{}^{\bullet+} \rightarrow CH_2Cl^+$ exhibits a Cl-KIE > 1, then the further dechlorination reaction $CH_2Cl^+ \rightarrow CH_2{}^+$ should present a Cl-KIE higher than 1 also. Since the primary Cl-KIE was measured as the MS signal intensity ratio of $CH_2{}^{35}Cl^{37}Cl^{\bullet+} \rightarrow CH_2{}^{37}Cl^+$ to $CH_2{}^{35}Cl^{37}Cl^{\bullet+} \rightarrow CH_2{}^{35}Cl^+$, the abundance of $CH_2{}^{37}Cl^+$ should be higher than that of $CH_2{}^{35}Cl^+$. When taking the further dechlorination into account, more amount of $CH_2{}^{35}Cl^+$ is further dechlorinated than that of $CH_2{}^{37}Cl^+$. Therefore, the measured Cl-KIEs should be higher than the real values during specific dechlorination reactions, which means the further dechlorination reactions fortify the measured Cl-KIEs. As a result, in this study the measured Cl-KIEs (mainly under unity) cannot be caused by the further Cl-KIEs that are higher than unity, and thus the measured Cl-KIEs lower than unity were really existent.

From the other side, we hypothesize that the primary and the further inter-ion Cl-KIEs were > 1, which is anticipative in accordance with the quasi-equilibrium theory (QET) and ZPE theory. The inter-ion Cl-KIEs were derived from the reactions of $CH_2{}^{35}Cl_2{}^{\bullet+} \rightarrow CH_2{}^{35}Cl^+$ and $CH_2{}^{37}Cl_2{}^{\bullet+} \rightarrow CH_2{}^{37}Cl^+$. We postulate that the amounts of the ions $CH_2{}^{35}Cl_2{}^{\bullet+}$ and $CH_2{}^{37}Cl_2{}^{\bullet+}$ are identical, then the abundance of the product ion $CH_2{}^{35}Cl^+$ prior to further dechlorination is higher than that of $CH_2{}^{37}Cl^+$. While $CH_2{}^{35}Cl^+$ is more likely to lose $^{35}Cl$ than $CH_2{}^{37}Cl^+$ to lose $^{37}Cl$ according to the hypothesis. Therefore, the primary and the further inter-ion Cl-KIEs make contribution to the measured Cl-KIEs for individual specific dechlorination reactions from opposite directions. In this context, it is possible that the measured Cl-KIEs can be either higher or lower than unity, depending on the magnitudes of the primary and the further inter-ion Cl-KIEs and the extents of dechlorination reactions. According to the QET and ZPE theory, low internal energy can enhance KIEs. However, the magnitudes (deviations from unity) of measured inter-ion Cl-KIEs (under unity) were increasing as the CEs increased from 6 to 60 eV. This observation indicates that the measured inter-ion Cl-KIEs were not likely triggered by the further Cl-KIEs in dissociation of product ions, particularly for those measured in the high energy region.

### 3.3.2. Isotope-competitive dechlorination reactions

Normal KIEs (>1) in MS are relatively easy to explain, since many studies have reported relevant observations and theoretical interpretations [14,32]. The normal KIEs can be successfully explained by the QET and ZPE theory. The critical energy ($E_0$) of an ion involving a bond with a light isotope is lower than that of the ion with a heavy isotope (note, critical energy is defined as the ZPE difference between a ground-state ion and its transition state [33]). Thus lighter isotopomer loses the light isotope more quickly than the heavier isotopomer to lose the heavy isotope, presenting normal KIE (>1). When internal energies of ions are near the critical energies of fragmentation reactions involving isotope atoms, the KIEs can be staggeringly large [1]. The measured normal Cl-KIEs of DCM in this study were in this scenario, showing decreasing tendencies from relatively large normal Cl-KIEs (1.020±0.001 and 1.026±0.002) to approximate unity with the CEs from 0 to 5 eV (Figure 3).

However, inverse KIEs (<1) occurring in MS are very challenging to interpret, and only very limited studies have reported the observations and putative mechanisms of inverse KIEs [14,19]. Green et al. studied the inverse Cl-KIEs of chlorine adducts of three simple chloroalkanes in CID by losing Cl$^-$ in MS/MS, and found that the Cl-KIE of chloroform/Cl$^-$ was extremely large and that of DCM/Cl$^-$ was 0.90±0.05 [14]. The authors concluded that the observed inverse Cl-KIEs during CID were attributable to the near-threshold dissociation of rotationally excited complexes. However, this near-threshold centrifugal effect cannot explain the fundamental mechanisms of the observed Cl-KIEs observed in our study, possibly due to that the cleaved C-Cl bonds are strong covalent bonds which are less sensitive to angular momentum effects in comparison with loosely bounded systems such as chloroform/Cl$^-$. In the study of Green et al., the Cl-KIEs strongly relied on the particular isotopomers and showed increasing trends as CEs increased [14]. To the contrary, in our study, the Cl-KIEs were all isotopomer-independent and exhibited decreasing tendencies as the CEs increased (Figure 3 and Figure 4). These differences may be also due to the different properties of the broken strong covalent bonds (C-Cl) and the dissociated loose coordinate bonds (C•••Cl$^-$).

The most notable phenomenon in our study is that the Cl-KIEs of DCM changed from large normal Cl-KIEs to extremely large inverse Cl-KIEs as the CEs varied from 0 to 60 eV, which has never been reported previously and is challenging to explain. To interpret this phenomenon, we refer to the theory in terms of competitive dechlorination reactions in EI-MS [34,35]. We take the reactions of $CH_2{}^{35}Cl^{37}Cl^{\bullet+} \rightarrow CH_2{}^{35}Cl^+$ (H) and $CH_2{}^{35}Cl^{37}Cl^{\bullet+} \rightarrow CH_2{}^{37}Cl^+$ (L) of DCM for example, which can be regarded as two isotope-competitive reactions during CID in MS/MS. As illustrated in Figure 6, the critical energy of reaction (H) is higher than that of reaction (L). Similarly, the internal energy threshold to generate just detectable product ion for the reaction (H) is higher than that for the reaction (L). At the internal energy of $E_1$, the rate constants of the two reactions reach equilibrium, i.e., $k_{35Cl} = k_{37Cl}$ or Cl-KIE = 1. In the lower internal energy region ($E_{(L)}$-$E_1$), the rate constant of reaction (L) is higher than that of reaction (H), showing the Cl-KIE higher than unity. Due to that the function of rate constant vs.

internal energy ($k(E)$) increases faster for the reaction (H) than for the reaction (L), the Cl-KIE gradually decreases to reach unity in the lower internal energy region, and then begin to be under unity as the internal energy continues to increase in the higher internal energy region (>$E_1$). This type of competitive reactions is always related to the looseness and tightness of transition states [36], and the $k(E)$ of the loose transition state increases faster than that of the tight [35]. A loose transition state means that the interacting fragments possess almost free rotations, while a tight transition state has the fragments whose rotations are to some extent hindered [37]. This indicates that the loose transition state has lower rotational barrier than the tight. Therefore, in the lower internal energy region the reaction rate constants are dominated by critical energies, while in the higher internal energy region the rate constants are controlled by rotational barriers. In this study, the transition state involving C--$^{37}$Cl may be slightly looser than that involving C--$^{35}$Cl, and therefore the Cl-KIE can decrease from higher than unity to lower than unity. Accordingly, the phenomenon of normal and inverse Cl-KIEs observed in this study can be interpreted. However, more fundamental mechanisms involving atomic and electronic motions for the phenomenon need further in-depth exploration.

In this study, we discovered that the Cl-KIE ranges of DCM were evidently larger than those of TCE and PCE, which suggests that the Cl-KIE magnitudes of DCM were significantly higher than those of TCE and PCE at the same CEs (Figure 3 and Figure 4). We infer that these differences might be owing to the C-Cl bond strength differences between DCM and the two chloroethylenes, given that the C-Cl bond strength of DCM (bond dissociation energy (BDE): 338.0±3.3 kJ/mol) is lower than those of the two chloroethylenes (BDEs ≥ 383.7 kJ/mol) [38]. In addition, as shown in Figure 4, the Cl-KIEs derived from PCE by losing two Cl atoms were significantly higher than that from PCE by losing one atom. We deduce that the available internal energy (AIE) for breaking one C-Cl bond was higher than the mean of the AIEs for breaking two C-Cl bonds during CID, thus leading to lower Cl-KIEs during the cleavage of one C-Cl bond than during the cleavages of two bonds since the Cl-KIEs decreased with the increase of CEs. We thus conclude that the Cl-KIEs during CID may depend on critical energies, bond strengths, AIEs, and transition state looseness/tightness (or rotational barriers).

### 3.4. Implications to environmental and atmospheric chemistry

The three investigated chlorohydrocarbons are important volatile/semi-volatile chlorinated organic pollutants in the environment, especially in the atmosphere and underground water [21-26]. Collisions between chlorohydrocarbon molecules and high-energy charged particles (e.g., solar wind and lightning current) in the atmosphere may lead to dechlorination reactions together with Cl-KIEs. In addition, solar radiation and cosmic rays could also trigger decompositions of chlorohydrocarbons in the atmosphere, accompanied by Cl-KIEs. In light of the results in this study, these Cl-KIEs occurring in the atmosphere are deduced to be energy-dependent, and able to appear in both normal and inverse modes. In addition, the magnitudes of these Cl-KIEs may be very large, owing to the high-energy charged particles and rays. The Cl-KIEs are expected to impact the chlorine isotope ratios and isotopologue distributions of chlorohydrocarbons in the atmosphere. As a result, this study could provide clues to investigate the variations of chlorine isotope ratios and isotopologue distributions of organochlorines in the atmosphere.

Furthermore, photodegradation is a promising approach to degrade organochlorine pollutants [39], in which Cl-KIEs may also take place and thereby influence the chlorine isotope ratios and isotopologue distributions of original and product organochlorines.

The Cl-KIEs can be applied to revealing mechanisms of the potodechlorination reactions, and accordingly provide strategies to optimize the photodegradation treatments.

## 4. Conclusions

In this study, we have systematically investigated the Cl-KIEs of DCM, TCE and PCE in CID by GC-MS/MS. Unusually large inverse Cl-KIEs were observed for all the three organochlorines. Specifically for DCM, both intra-ion and inter-ion Cl-KIEs were explored, exhibiting both normal and inverse Cl-KIEs depending on CEs. The observed Cl-KIEs generally showed declining tendencies as the CEs increased from 0 to 60 eV, presenting evident CE-dependent features, but had no relationship with MS signal intensities, injection concentrations/volumes, and isotopologues' relative abundances. The observed Cl-KIEs were deduced to not be attributable to further dechlorination reactions of product ions, in which Cl-KIEs were anticipated to occur also. The mechanisms of normal Cl-KIEs can be explained by the QET and ZPE theory, and those of inverse Cl-KIEs are more challenging to elucidate and have been interpreted by isotope-competitive reactions related to transition state theory. In a lower internal energy region, the Cl-KIEs are inferred to be dominated by critical energies, while in a higher internal energy region they are deduced to be controlled by looseness/tightness of transition states (or rotational barriers). Overall, the Cl-KIEs may be dependent on critical energies, bond strengths, AIEs, and looseness/tightness of transition states. This study may help to explore the changes of chlorine isotope ratios and isotopologue distributions of organochlorines in the atmosphere, in view of that DCM, TCE and PCE are important volatile/semi-volatile organochlorine pollutants in the environment, particularly in the atmosphere. The findings of this study could be extended to revelation of Cl-KIEs in potodechlorination reactions along with reaction mechanisms, and thereby providing strategies to achieve optimal photodegradation parameters.

# Appendix A. Supplementary data

The Supplementary data is available on the website at http://pending.

# Acknowledgements

This study was financially supported by the National Natural Science Foundation of China (Grant No. 41603092).

## Figure legends

**Figure 1**. Representative chromatograms of the multiple reaction monitoring (MRM) transitions of $m/z$ 86.0→49.0 and 86.0→51.0 of dichloromethane (DCM) at different collision energies (CEs) by GC-MS/MS. The discrepancies between the two MRM transitions point to the intra-ion chlorine kinetic isotope effects (Cl-KIEs) of DCM.

**Figure 2**. Intra-ion and inter-ion Cl-KIEs of DCM measured with the combined scan and separated scans with and without background subtraction. Combined scan: monitoring all MRM transitions of DCM; separated scans: monitoring merely $m/z$ 86.0→49.0 and 86.0→51.0, or only $m/z$ 84.0→49.0 and 88.0→51.0. The CE was set at 45 eV. Error bars represent the standard deviations (1σ, n = 6, the same below).

**Figure 3**. Measured Cl-KIEs of DCM during collision-induced dissociation (CID) at different CEs. The CEs are divided into two regions, i.e., a low energy region (0-5 eV) and a high energy region (6-60 eV). Note, (a) and (b): Cl-KIEs measured at the low and the high energy regions, respectively; $k_{35Cl}$ and $k_{37Cl}$ indicate the rate constants corresponding to the dechlorination reactions by losing $^{35}$Cl and $^{37}$Cl atoms, respectively; solid curves represent exponential regressions and shaded areas indicate the corresponding 95% confidence intervals (similarly hereinafter); the fitted functions for intra-ion Cl-KIEs vs. CEs and inter-ion Cl-KIEs vs. CEs in (a) were y = 0.986 + 0.033exp(−0.349x) ($R^2$ = 0.995) and y = 0.990 + 0.034exp(−0.211x) ($R^2$ = 0.959), respectively, and those in (b) were y = 1.117 − 0.094exp(0.033x) ($R^2$ = 0.983) and y = 1.014 − 0.021exp(0.050x) ($R^2$ = 0.989), respectively; the decimals of the fitting parameters were displayed according to the respective standard deviations (1σ, similarly hereinafter).

**Figure 4**. Measured Cl-KIEs of trichloroethylene (TCE) and tetrachloroethylene (PCE) during CID at different CEs. (a): the Cl-KIEs of TCE correspond to the reactions of $m/z$ 131.9→94.9 vs. 131.9→96.9, and $m/z$ 133.9→96.9 vs. 133.9→98.9; (b): the Cl-KIEs of PCE correspond to the reactions of $m/z$ 165.9→128.9 vs. 165.9→130.9, and $m/z$ 167.9→130.9 vs. 167.9→132.9; (c): the Cl-KIEs of PCE correspond to the reactions of $m/z$ 167.9→130.9 vs. 167.9→132.9, and $m/z$ 167.9→93.9 vs. 167.9→97.9 in the CE region of 20-60 eV; solid lines denote linear regressions; the fitted functions for Cl-KIEs vs. CEs of the reactions of $m/z$ 131.9→94.9 vs. 131.9→96.9, $m/z$ 133.9→96.9 vs. 133.9→98.9, $m/z$ 165.9→128.9 vs. 165.9→130.9, $m/z$ 167.9→130.9 vs. 167.9→132.9, $m/z$ 167.9→130.9 vs. 167.9→132.9, and $m/z$ 167.9→93.9 vs. 167.9→97.9 were y = 0.999 − 0.001x ($R^2$ = 0.996), y = 0.990 − 0.001x ($R^2$ = 0.979), y = 0.983 − 0.001x ($R^2$ = 0.990), y = 0.977 − 0.001x ($R^2$ = 0.948), y = 0.990 − 0.002x ($R^2$ = 0.963), and y = 1.036 − 0.001x ($R^2$ = 0.958), respectively.

**Figure 5**. Correlations between Cl-KIEs and MS signal intensities of DCM obtained at different CEs. Note, (a): correlations at the low energy region (0-5 eV); (b): correlations at the high energy region (6-60 eV). ∑Intensity for intra-ion Cl-KIEs is the sum of the MS signal intensities of $m/z$ 86.0→49.0 and 86.0→51.0; ∑Intensity for inter-ion Cl-KIEs is the sum of the MS signal intensities of $m/z$ 84.0→49.0 and 88.0→51.0; the fitted functions for intra-ion Cl-KIEs vs. ∑Intensities and inter-ion Cl-KIEs vs. ∑Intensities in (a) were y = 1.040 − 2.7×10$^{-8}$x ($R^2$ = 0.965) and y = 1.039 − 1.2×10$^{-8}$x ($R^2$ = 0.932), respectively and those in (b) were y = 1.031 − 6.43×10$^4$exp(−8.8×10$^{-6}$x) ($R^2$ = 0.999) and y = 0.977 − 0.430exp(−9.8×10$^{-6}$x) ($R^2$ = 0.996), respectively.

**Figure 6**. Illustration of the correlations between reaction rate constants (log k(E)) and internal energy (E) of two isotope-competitive dechlorination reactions during CID by GC-EI-MS/MS. Note, $E_{0(L)}$ and $E_{0(H)}$ denote the critical energies for the reactions $m/z$ 86.0→51.0 (L) and $m/z$ 86.0→49.0 (H), respectively; $E_{(L)}$ and $E_{(H)}$ correspond to the energy thresholds generating detectable ions of $m/z$ 51.0 and 49.0, respectively; $E^*_{(L)}$ and $E^*_{(H)}$ indicate the kinetic shifts for the reactions giving rise to detectable ions of $m/z$ 51.0 and 49.0, respectively; -log $\tau$ represents the rate constant threshold where the product ions can just start to be detectable; $E_1$ is the internal energy of the precursor ion ($m/z$ 86.0), at which the rate constants of the two reactions are equal. This diagram is referring to the literature [35].

# Figures

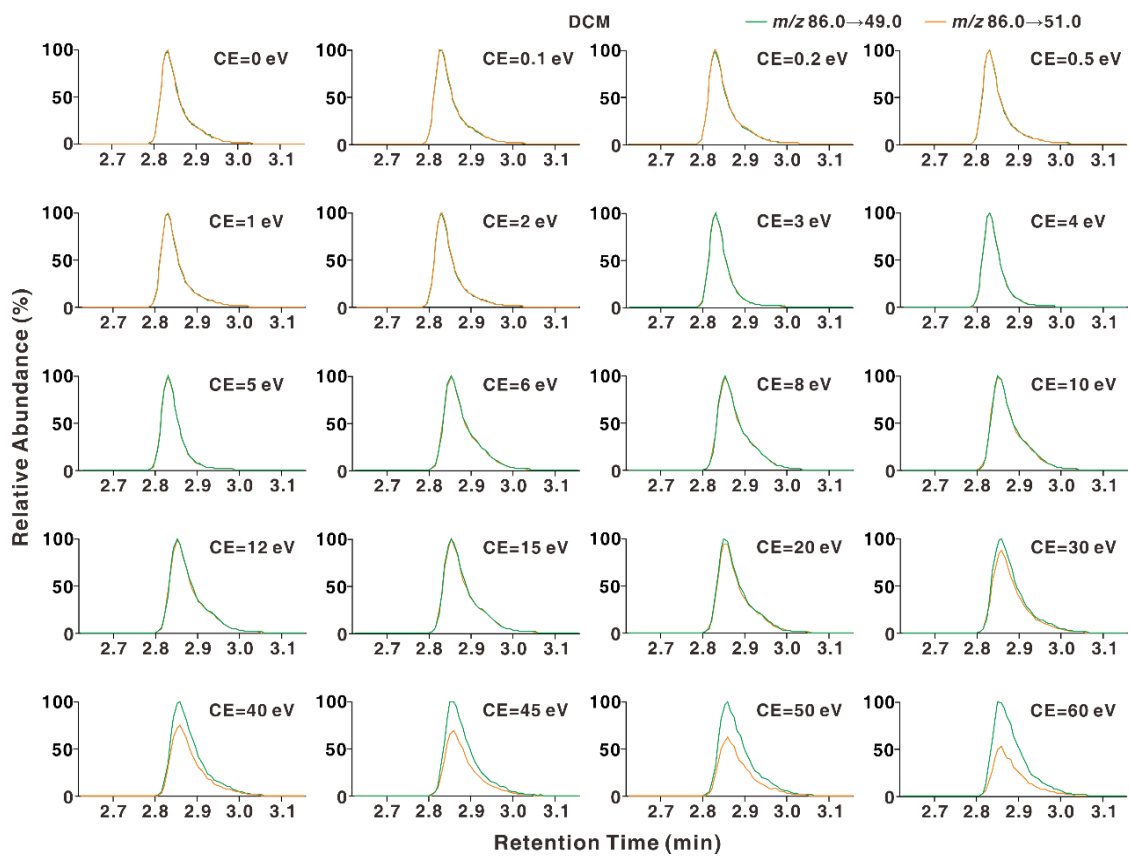

**Figure 1**. Representative chromatograms of the multiple reaction monitoring (MRM) transitions of $m/z$ 86.0→49.0 and 86.0→51.0 of dichloromethane (DCM) at different collision energies (CEs) on GC-MS/MS. The discrepancies between the two MRM transitions point to the intra-ion chlorine kinetic isotope effects (Cl-KIEs) of DCM.

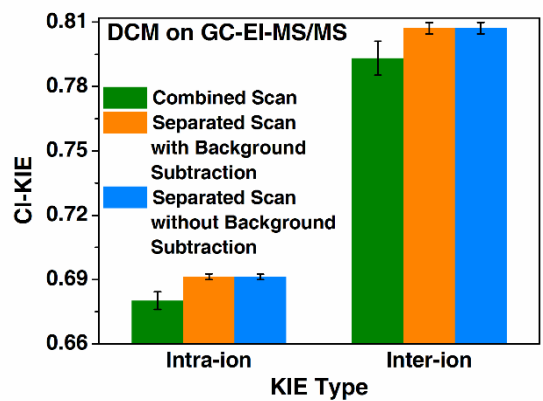

**Figure 2**. Intra-ion and inter-ion Cl-KIEs of DCM measured with the combined scan and separated scans with and without background subtraction. Combined scan: monitoring all MRM transitions of DCM; separated scans: monitoring merely *m/z* 86.0→49.0 and 86.0→51.0, or only *m/z* 84.0→49.0 and 88.0→51.0. The CE was set at 45 eV. Error bars represent the standard deviations (1σ, n = 6, the same below).

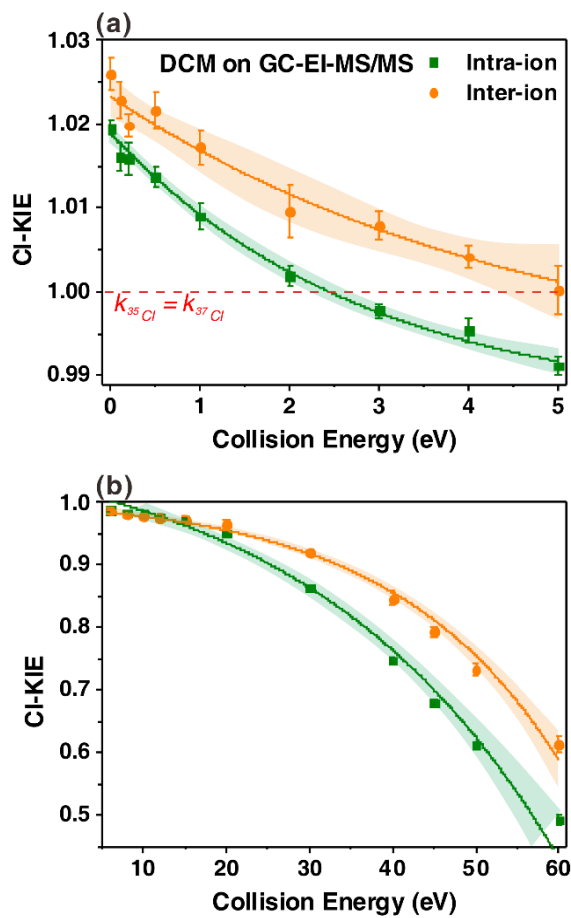

**Figure 3**. Measured Cl-KIEs of DCM during collision-induced dissociation (CID) at different CEs. The CEs are divided into two regions, i.e., a low energy region (0-5 eV) and a high energy region (6-60 eV). Note, (a) and (b): Cl-KIEs measured at the low and the high energy regions, respectively; $k_{35Cl}$ and $k_{37Cl}$ indicate the rate constants corresponding to the dechlorination reactions by losing $^{35}$Cl and $^{37}$Cl atoms, respectively; solid curves represent exponential regressions and shaded areas indicate the corresponding 95% confidence intervals (similarly hereinafter); the fitted functions for intra-ion Cl-KIEs vs. CEs and inter-ion Cl-KIEs vs. CEs in (a) were y = 0.986 + 0.033exp(−0.349x) ($R^2$ = 0.995) and y = 0.990 + 0.034exp(−0.211x) ($R^2$ = 0.959), respectively, and those in (b) were y = 1.117 − 0.094exp(0.033x) ($R^2$ = 0.983) and y = 1.014 − 0.021exp(0.050x) ($R^2$ = 0.989), respectively; the decimals of the fitting parameters were displayed according to the respective standard deviations (1σ, similarly hereinafter).

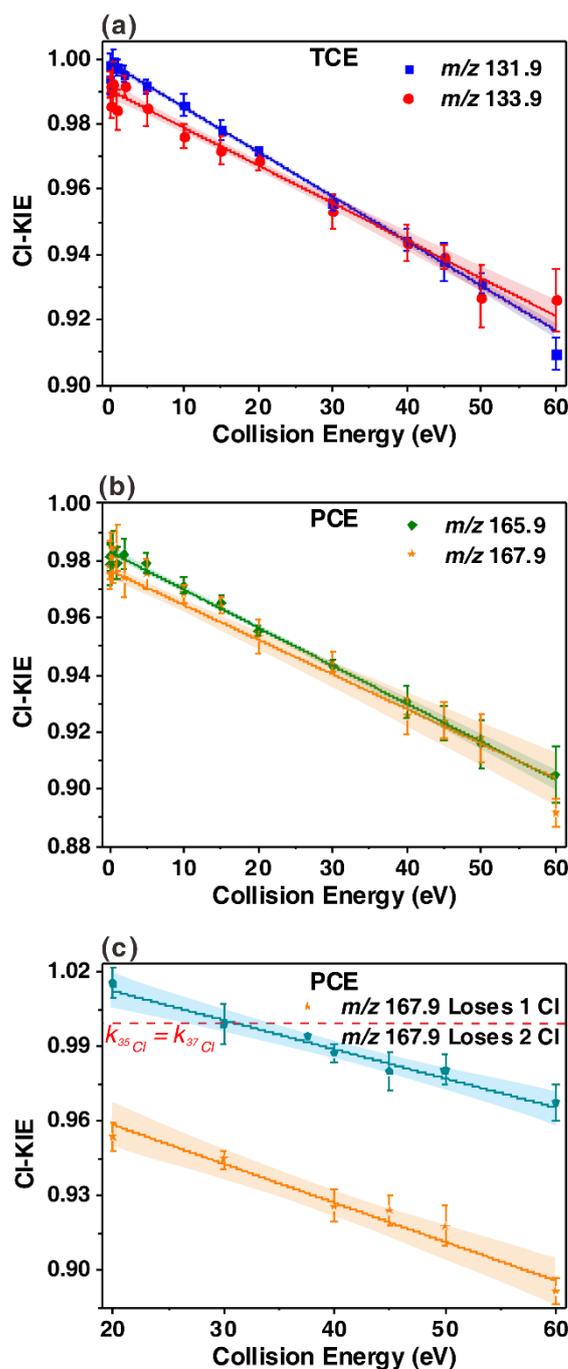

**Figure 4**. Measured Cl-KIEs of trichloroethylene (TCE) and tetrachloroethylene (PCE) during CID at different CEs. (a): the Cl-KIEs of TCE correspond to the reactions of $m/z$ 131.9→94.9 vs. 131.9→96.9, and $m/z$ 133.9→96.9 vs. 133.9→98.9; (b): the Cl-KIEs of PCE correspond to the reactions of $m/z$ 165.9→128.9 vs. 165.9→130.9, and $m/z$ 167.9→130.9 vs. 167.9→132.9; (c): the Cl-KIEs of PCE correspond to the reactions of $m/z$ 167.9→130.9 vs. 167.9→132.9, and $m/z$ 167.9→93.9 vs. 167.9→97.9 in the CE region of 20-60 eV; solid lines denote linear regressions; the fitted functions for Cl-KIEs vs. CEs of the reactions of $m/z$ 131.9→94.9 vs. 131.9→96.9, $m/z$ 133.9→96.9 vs. 133.9→98.9, $m/z$ 165.9→128.9 vs. 165.9→130.9, $m/z$ 167.9→130.9 vs. 167.9→132.9, $m/z$ 167.9→130.9 vs. 167.9→132.9, and $m/z$ 167.9→93.9 vs. 167.9→97.9 were y = 0.999 − 0.001x ($R^2$ = 0.996), y = 0.990 − 0.001x ($R^2$ = 0.979), y = 0.983 − 0.001x ($R^2$ = 0.990), y = 0.977 − 0.001x ($R^2$ = 0.948), y = 0.990 − 0.002x ($R^2$ = 0.963), and y = 1.036 − 0.001x ($R^2$ = 0.958), respectively.

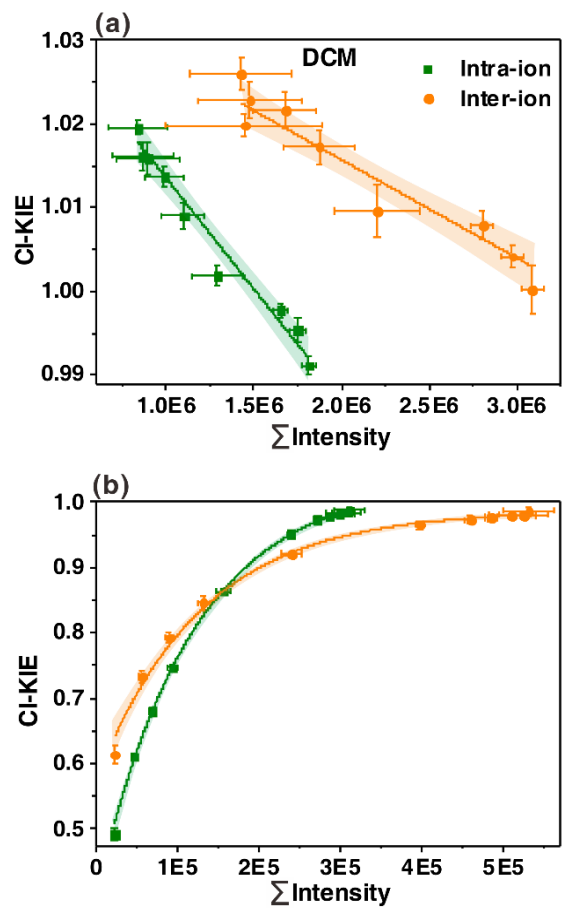

**Figure 5**. Correlations between Cl-KIEs and MS signal intensities of DCM obtained at different CEs. Note, (a): correlations at the low energy region (0-5 eV); (b): correlations at the high energy region (6-60 eV). $\sum$Intensity for intra-ion Cl-KIEs is the sum of the MS signal intensities of $m/z$ 86.0→49.0 and 86.0→51.0; $\sum$Intensity for inter-ion Cl-KIEs is the sum of the MS signal intensities of $m/z$ 84.0→49.0 and 88.0→51.0; the fitted functions for intra-ion Cl-KIEs vs. $\sum$Intensities and inter-ion Cl-KIEs vs. $\sum$Intensities in (a) were y = $1.040 - 2.7\times10^{-8}$x ($R^2 = 0.965$) and y = $1.039 - 1.2\times10^{-8}$x ($R^2 = 0.932$), respectively and those in (b) were y = $1.031 - 6.43\times10^4\exp(-8.8\times10^{-6}$x) ($R^2 = 0.999$) and y = $0.977 - 0.430\exp(-9.8\times10^{-6}$x) ($R^2 = 0.996$), respectively.

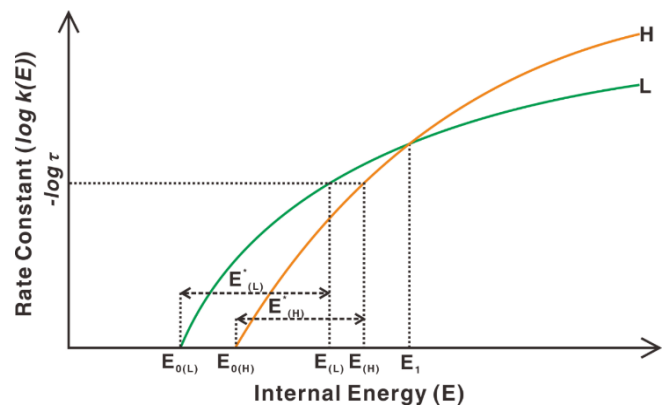

**Figure 6**. Illustration of the correlations between reaction rate constants (log k(E)) and internal energy (E) of two isotope-competitive dechlorination reactions during CID on GC-EI-MS/MS. Note, $E_{0(L)}$ and $E_{0(H)}$ denote the critical energies for the reactions $m/z$ 86.0→51.0 (L) and $m/z$ 86.0→49.0 (H), respectively; $E_{(L)}$ and $E_{(H)}$ correspond to the energy thresholds generating detectable ions of $m/z$ 51.0 and 49.0, respectively; $E^{*}_{(L)}$ and $E^{*}_{(H)}$ indicate the kinetic shifts for the reactions giving rise to detectable ions of $m/z$ 51.0 and 49.0, respectively; -log $\tau$ represents the rate constant threshold where the product ions can just start to be detectable; $E_1$ is the internal energy of the precursor ion ($m/z$ 86.0), at which the rate constants of the two reactions are equal. This diagram is referring to the literature [35].